\documentclass[apj]{emulateapj}
\usepackage[colorlinks,linkcolor={blue},citecolor={blue},urlcolor={blue}]{hyperref}
\usepackage{threeparttable}
\usepackage{bm}
\usepackage{graphicx}
\usepackage{epsf}
\usepackage{graphics}
\usepackage{amsmath}
\usepackage{lineno}
\usepackage{multirow,booktabs}
\usepackage{savesym}
\savesymbol{tablenum}
\usepackage{siunitx}
\restoresymbol{SIX}{tablenum}

\def\beq{\begin{equation}}
\def\eeq{\end{equation}}
\def\bey{\begin{eqnarray}}
\def\eey{\end{eqnarray}}

\def\Msun{{\rm M_\odot}}

\def\gs{\mathrel{\raise1.16pt\hbox{$>$}\kern-7.0pt
\lower3.06pt\hbox{{$\scriptstyle \sim$}}}}
\def\ls{\mathrel{\raise1.16pt\hbox{$<$}\kern-7.0pt
\lower3.06pt\hbox{{$\scriptstyle \sim$}}}}
\def\gtsima{\, {\buildrel > \over \sim} \,}
\def\ltsima{\, {\buildrel < \over \sim} \,}
\def\prosima{\, {\buildrel \propto \over \sim} \,}
\def\gsim{\lower.5ex\hbox{\gtsima}}
\def\lsim{\lower.5ex\hbox{\ltsima}}
\def\simgt{\lower.5ex\hbox{\gtsima}}
\def\simlt{\lower.5ex\hbox{\ltsima}}
\def\simpr{\lower.5ex\hbox{\prosima}}

\shorttitle{Early-type to Late-type}
\shortauthors{M., Zhou et al.}

\begin{document}
\title{Evidence for the transformation from lenticular to spiral galaxies}

\author{Mengkui Zhou\altaffilmark{1,2}, Huiyuan Wang\altaffilmark{1,2}, Ran Li\altaffilmark{3}, 
Yangyao Chen\altaffilmark{4,5}, Hui Hong\altaffilmark{6}, Houjun Mo\altaffilmark{7}, Yu Rong\altaffilmark{1,2}, Enci Wang\altaffilmark{1,2}, Huiling liu\altaffilmark{1,2}, Zhicheng He\altaffilmark{1,2}, Ziwen Zhang\altaffilmark{1,2}}
\altaffiltext{1}{Department of Astronomy, University of Science and Technology of China, Hefei, Anhui 230026, China;  whywang@ustc.edu.cn} 
\altaffiltext{2}{School of Astronomy and Space Science, University of Science and Technology of China, Hefei 230026, China}
\altaffiltext{3}{School of Physics and Astronomy, Beijing Normal University,  Beijing 100875, China}
\altaffiltext{4}{School of Astronomy and Space Science, Nanjing University, Nanjing, Jiangsu 210093, China}
\altaffiltext{5}{Key Laboratory of Modern Astronomy and Astrophysics, Nanjing University, Ministry of Education, Nanjing, Jiangsu 210093, China}
\altaffiltext{6}{Department of Physics \& Astronomy, University of California, Riverside, CA, 92521, USA}
\altaffiltext{7}{Department of Astronomy, University of Massachusetts, Amherst MA 01003-9305, USA}

\begin{abstract}
It is widely accepted that late-type galaxies, such as spirals, evolve into early-type systems, including elliptical and lenticular galaxies, through galaxy mergers and violent disk instability processes. Throughout this morphological transformation, star formation is typically suppressed by quenching mechanisms whose detailed nature remains the subject of active investigation. Here, we present compelling evidence for an evolutionary pathway that proceeds in the reverse direction. Using the integral field unit observations, we identify a population of spiral galaxies hosting quenched central cores (QCCs). These galaxies exhibit bimodal distributions in both their stellar population properties and their dynamical properties, along with sharp changes in radial gradients near the QCC boundary. These results indicate that the QCCs and the surrounding outer disks formed at distinct cosmic epochs and through different physical processes. Remarkably, QCCs closely resemble quiescent early-type galaxies, particularly lenticular galaxies, in their mass-size and mass-velocity dispersion scaling relations, as well as in their stellar population demographics and internal kinematics. These findings provide strong support for a rejuvenation scenario in which spiral disks are reassembled around pre-existing quiescent lenticular or early-type systems. Moreover, we show that such rejuvenation, accompanied by a reverse morphological transformation from early- to late-type appearance, is quite common. This indicates that quenching in galaxies is not invariably a terminal state and can be reversed under appropriate conditions.
\end{abstract}


\section{Introduction} \label{sec_intro}

The Hubble sequence, originally formulated a century ago (\citealt{Hubble_1926}), continues to serve as the standard framework for the morphological classification of galaxies, categorizing systems into elliptical, lenticular (S0), spiral and irregular galaxies. A central feature of this scheme is the dichotomy between ``early-type'' (elliptical and lenticular) and ``late-type'' (spiral and irregular) galaxies. Generally, late-type galaxies (LTGs) exhibit extended disk structures and blue optical colors, indicative of ongoing star formation. In contrast, early-type galaxies (ETGs) typically exhibit compact morphologies and red stellar populations with minimal or even undetectable star-formation activity. As a result, we can see clear bimodality in the distributions of both galaxy morphology and star-formation activity in the local Universe and higher redshift \citep{Strateva2001AJ,Baldry2004ApJ,Bell2012ApJ,Barro2017ApJ, Wang_2018a}.

Observations indicate that the fraction of quiescent galaxies increases steadily over cosmic time (e.g., \citealt{Ilbert_2013}; \citealt{Muzzin_2013}; \citealt{Davidzon_2017}; \citealt{Weaver_2022}), implying a primary evolutionary pathway in which star-forming systems are transformed into quenched galaxies. Investigating the nature of this quenching process is therefore crucial to our understanding of galaxy evolution, and a wide range of physical mechanisms have been proposed to explain it (e.g., \citealt{Gunn_1972}; \citealt{Birnboim_2003}; \citealt{Croton_2006}; \citealt{Dekel_2006}; \citealt{Martig_2009}; \citealt{Peng_2010}; \citealt{Fabian_2012}; \citealt{heckman_2014}; \citealt{WangH2018}; \citealt{Bluck_2020b}; \citealt{Gensior_2020}). Furthermore, previous studies have demonstrated that quenching is strongly correlated with the bulge-to-total mass ratio, central surface mass density, and stellar velocity dispersion (\citealt{FangJ2013ApJ}; \citealt{Barro2017ApJ}; \citealt{WangE2018ApJ}; \citealt{Bluck_2020b}; \citealt{hongDynamicalHotnessStar2023}), implying that the cessation of star formation is associated with galaxy morphology. Because most star-forming galaxies have late-type morphologies, whereas the quenched population is dominated by early-types, an evolutionary sequence from late-type to early-type galaxies emerges as a natural scenario.

Extensive studies have indeed confirmed that a morphological transformation from late-type to early-type galaxies is feasible. Numerical simulations, for example, indicate that major mergers or interactions can efficiently disrupt the spiral and disk components of progenitor galaxies, thus producing elliptical systems \citep{White1978MNRAS,Springel2005ApJ,Cox2006ApJ,WangY2026}. These theoretical results are supported by high-resolution observations of nearby ellipticals, which frequently exhibit tidal features and other merger signatures \citep[e.g.][]{Tal2009AJ}. In addition, such transformations can also occur through internal mechanisms, including violent disk instabilities, which can redistribute galactic material and build up spheroidal components at the center of galaxies \citep[e.g.][]{Kormendy2004ARA&A}.

Nevertheless, the reality is likely more complicated. The detection of recent star formation in early-type galaxies (e.g., \citealt{Kaviraj_2007}; \citealt{Salim2012ApJ}; \citealt{Paudel_2023}) indicates that star formation can resume in galaxies that were previously quenched. Specifically, evidence indicates that gas accretion or capture of gas-rich satellites can supply the material required to trigger such rejuvenation events (e.g. \citealt{Rathore_2022}; \citealt{WangYj_2025}). In addition, progress in advanced SED-fitting methods now allows for detailed reconstructions of star formation histories, making it possible to identify rejuvenated galaxies within larger samples (e.g., \citealt{Chauke_2019}; \citealt{Woodrum_2022}; \citealt{Tanaka_2024}). Collectively, these works imply that the transition from a quenched phase back to an actively star-forming phase occurs at a non-negligible rate (e.g., \citealt{Tanaka_2024}). However, the rejuvenation episodes discussed in these works are generally modest in extent and do not suffice to induce major morphological transformations. 

Some studies propose that quenched early-type galaxies can rebuild their stellar disks through gas accretion or gas-rich minor mergers (e.g., \citealt{Graham_2015}; \citealt{Hon_2022}; \citealt{Yun2026}). \cite{Hon_2022} found that the volume number density of local elliptical galaxies is significantly lower than the peak number density of massive red nuggets at $z\sim1-2$, but that of local compact spheroids, obtained via multi-component decomposition, is comparable. They argued that the massive red nuggets have been cloaked by newly-formed surrounding disks and now serve as the compact central bulges of S0 or spiral galaxies. Nevertheless, an important caveat of their work is that it is confined to galaxies within 110 Mpc, a volume dominated by the local void, where the number density of massive galaxies is significantly underestimated due to cosmic variance (see \citealt{ChenY2019ApJ}). In addition, these studies did not investigate the similarity or difference between the bulges and the red nuggets. More recently, \cite{Yun2026} reported that, in an interacting pair, an elliptical galaxy is developing a rotating disk by accreting gas from its companion galaxy.

Despite the uncertainties in observational studies, the physical feasibility of this structural regrowth scenario is increasingly supported by numerical simulations and theoretical models. Specifically, \citet{Diaz_2018} utilized hydrodynamic simulations to show that elliptical galaxies can merge with gas-rich spirals to evolve into S0 systems. Additional support for this scenario comes from \citet{Sparre&Springel_2017}, who showed that remnants of major mergers can rebuild prominent stellar disks and evolve back into active spiral systems. Theoretically, this "inside-out" assembly is consistent with the two-phase formation models discussed in \citet{Driver2013MNRAS} and \citet{Mo_2024}. In such a framework, a rapidly formed, dynamically hot spheroid emerges during an initial fast phase of high gas fraction, while a rotationally supported cold disk develops during a subsequent, slow phase as the potential stabilizes \citep{Mo_2024}. 

So, a comprehensive analysis of observational data with a different approach is clearly needed. The Sloan Digital Sky Survey (SDSS) Mapping Nearby Galaxies at Apache Point Observatory (MaNGA; \citealt{Bundy_2015}) project delivers spatially resolved spectroscopic data for approximately 10,000 galaxies within a redshift range $0.01<z<0.15$, thereby providing an exceptional resource for investigating star formation quenching and its association with galaxy morphology. This dataset enables the extraction of spatially resolved quenching information, facilitating  classification of galaxies and identification of systems within galaxies based on spatially resolved spectroscopic information. Numerous studies have utilized MaNGA data to compare radial profiles across different galaxy types (e.g., \citealt{Li_2015}; \citealt{Zheng_2017}; \citealt{Parikh_2021}).

This paper is organized as follows. In Section \ref{sec_obs}, we describe the samples used in this paper and the method for identifying the quenched central cores (QCCs). Section \ref{sec_bimod} analyzes and presents the bimodality and the special radial profiles in individual spiral galaxies with QCCs. Section \ref{sec_s&d} provides a detailed comparison between QCCs and quenched early-type galaxies, with a particular focus on the scaling relations and the radial profiles of their stellar properties. Finally, we discuss our findings and summarize our conclusions in Section \ref{sec_sum}. Throughout the paper, we assume the following $\Lambda$CDM cosmology parameters: $H_{\rm 0}=73\, \rm km/s/Mpc$, $\Omega_{\rm M}=0.3$, and $\Omega_{\rm \Lambda}=0.7$.

\section{Methodology}\label{sec_obs}


In this section, we introduce our galaxy sample and the methods to classify galaxies and analyze the spatially resolved data.

\subsection{MaNGA survey and sample selection}

Our galaxy sample is taken from MaNGA (\citealt{Bundy_2015}) of SDSS Data Release 17 (DR17, \citealt{Abdurrouf_2022}). MaNGA galaxies are observed by integral field units (IFUs), which have a wavelength coverage from 3,600 to 10,300 \AA~ and a spectral resolution of about 2000, giving an instrumental resolution of $\sim60~\mathrm{km \, s^{-1}}$, which are particularly beneficial for efficiently modeling stellar populations and their properties in general. The spatial resolution is 2.5 arcsec FWHM after combining the dithered images. The field of view (FOV) of the MaNGA survey covers an aperture of $1.5\,R_{\rm e}$ or larger, where $R_{\rm e}$ is the effective radius of the galaxy. This allows us to classify galaxies based on their spatially resolved quenching properties and to further analyze the spatially resolved characteristics of these galaxies. 

In this work, we utilize the data products derived from \texttt{pyPipe3D} (\citealt{2016RMxAA..52...21S, 2016RMxAA..52..171S, Sánchez_2022}). Pipe3D provides data for 10,220 galaxies derived from a comprehensive analysis of 10,245 datacubes. It performs an SSP (Simple Stellar Population) decomposition of the stellar continuum followed by emission-line analysis. To ensure sufficient signal-to-noise ratios, it employs a spatial binning (tessellation) scheme before fitting. The stellar population properties, including stellar age and metallicity ($Z_*$), are derived by fitting the continuum with the MaStar\_sLOG library (\citealt{Yan_2019}) consisting of 273 SSP templates with 39 ages (from $\rm 1~Myr$ to $\rm 13.5~Gyr$) and seven metallicities (from $0.006~Z_\odot$ to $2.353~Z_\odot$), assuming a Salpeter (\citealt{Salpeter_1955}) initial mass function (IMF). During this procedure, the SSP templates are shifted according to the mean stellar velocity ($v_*$), attenuated by the dust curve \citep{Cardelli_1989}, and Gaussian-broadened by the velocity dispersion ($\sigma_*$) to match the observed continuum spectra. 

The spatially resolved parameters, such as stellar age, stellar metallicity, $\sigma_*$, $v_*$, stellar surface mass density ($\Sigma_*$) and $4000\,\rmÅ$ break (D4000), are direct products of Pipe3D. 
Note that the stellar age and metallicity used in this paper are luminosity-weighted averages, with each SSP component weighted by its fractional flux contribution at the normalization wavelength.
Pipe3D also performs a series of Monte Carlo iterations by perturbing the input spectrum with its errors to estimate the uncertainties of the derived physical parameters. The stellar surface mass density is estimated based on the derived mass-to-light ratio and the local surface brightness. The $4000\,\rmÅ$ break is defined as the flux ratio between the red and blue sides of $\lambda4000\,\rmÅ$, as described in \cite{bruzual_a_spectral_1983} and \cite{Balogh_1999}.
To ensure the reliability  of these spatially resolved parameters throughout this study, we utilize the \texttt{GAIA\_MASK} and \texttt{SELECT\_REG} extensions provided by Pipe3D to remove spaxels with foreground starlight pollution or low S/N continuum spectra. 

Pipe3D also provides global parameters for each galaxy, for example, redshift, effective radius ($R_{\rm e}$), position angle (PA) of major axis and ellipticity ($e=\sqrt{1-(b/a)^2}$), which are directly taken from NSA catalog. In addition, the total stellar mass ($M_*$) and the $\rm SFR$ are obtained by integrating the corresponding values of spaxels within the FOV, excluding the masked spaxels. We subsequently excluded galaxies that do not possess valid NSA redshifts or $R_{\rm e}$, resulting in a sample of 9,988 galaxies.
Following \cite{hongDynamicalHotnessStar2023}, we also exclude galaxies with low S/N through the QCFLAG and galaxies without valid measurements for some parameters.
Moreover, to exclude the influence of environmental quenching mechanism, we cross-match our sample with the SDSS group catalog provided by \cite{Yang_2007}, and obtain 6,231 central galaxies. In this paper, we only consider massive galaxies.
Our galaxy sample consists of 5,124 galaxies with stellar masses covering the range $10^{10.0}-10^{12.0}\rm\,M_\odot$.

\subsection{Galaxy morphology}

For the morphology classification, we use the MaNGA morphology deep-learning DR17 catalog (\citealt{dominguez_sanchez_improving_2018, dominguezsanchez_sdss-iv_2021}). This catalog classifies galaxies in the final SDSS-MaNGA DR17 sample using a convolutional neural network (CNN) trained on visual classifications from both the \cite{Nair_2010} catalog of SDSS DR4 galaxies and Galaxy Zoo 2 (\citealt{willett_galaxy_2013}), which achieves $> 90\%$ classification accuracy. The catalog outputs include continuous T-Type classifications [-4, 9], binary ETG/LTG classifications with uncertainty estimates. These deep learning classifications complement traditional visual morphology catalogs by providing quantitative, reproducible measurements for the complete MaNGA sample. Moreover, all morphological classifications were visually verified to ensure robustness. The final catalog provides both the visual classification (VC) and a corresponding reliability flag (VF). Galaxies are assigned a VC code based on morphology: elliptical (E, VC=1), lenticular (S0, VC=2), and late-type galaxy (LTG, VC=3). To ensure data reliability, we first exclude all galaxies flagged as VF=1 from the analysis. Following the stringent selection criteria recommended by \cite{dominguezsanchez_sdss-iv_2021},  we then classify the remaining galaxies into Spirals, S0s, and Ellipticals:

\begin{itemize}
    \item Spiral: $\text{T-Type} > 0$ \textbf{AND} $\mathrm{P\_LTG} \geq 0.5$ \textbf{AND} $\mathrm{VC}=3$;
    \item S0: $\text{T-Type} \leq 0$ \textbf{AND} $\mathrm{P\_S0} > 0.5$ \textbf{AND} $\mathrm{P\_LTG} < 0.5$ \textbf{AND} $\mathrm{VC} = 2$;
    \item Elliptical: $\text{T-Type} \leq 0$ \textbf{AND} $\mathrm{P\_S0} \leq 0.5$ \textbf{AND} $\mathrm{P\_LTG} < 0.5$ \textbf{AND} $\mathrm{VC} = 1$,
\end{itemize}
where $\rm P\_{\text{LTG}}$ and $\rm P\_{\text{S0}}$ are the probabilities of being classified as an LTG and an S0, respectively. Based on the criteria defined above, the majority (4,435/5,124) of galaxies in our parent sample are successfully classified. 1,521 galaxies are classified as elliptical galaxies, 445 as S0 and 2,469 as spiral galaxies.

\begin{table*}[t]
    \centering
    \caption{Samples and abbreviations}\label{tab:abbrev}
    \begin{tabular}{ccc}
    \toprule
        Abbreviations & Description & Number\\
        \midrule
        QCCs (in spiral) & quenched cores in the central regions of spiral galaxies & 247\\
        &  within which 95\% of spaxels have D4000 $>1.55$ & \\
        ODKs & outer disk regions of spiral galaxies hosting QCCs & 247\\
        \midrule
        Spirals with QCC &  spiral galaxies hosting QCCs & 247 \\
        Spirals with bimodality &  spirals hosting QCCs and showing bimodality in both D4000 and $\sigma_*$ & 154 \\
        Spirals without bimodality & spirals hosting QCCs and showing no bimodality in D4000 or $\sigma_*$ & 93\\
        \midrule
        Quenched ellipitcals & elliptical galaxies with more than 95\% of spaxels within $1.5\,R_{\rm e}$ having D4000 $>1.55$ & 1097 \\
        Quenched S0s & lenticular galaxies with more than 95\% of spaxels within $1.5\,R_{\rm e}$ having D4000 $>1.55$ & 264\\
        Quenched early-type & collection of quenched ellpiticals and quenched S0s & 1361\\
        \midrule
        Star-forming spiral & spiral galaxies with more than 95\% of spaxels within $1.5\,R_{\rm e}$ having D4000 $<1.55$ & 1561\\
        \bottomrule
    \end{tabular}
\end{table*}

\subsection{Quenched central cores in spiral galaxies} \label{subsec_QCCdef}

\begin{figure*}[t]
    \centering
    \includegraphics[width=0.7\linewidth]{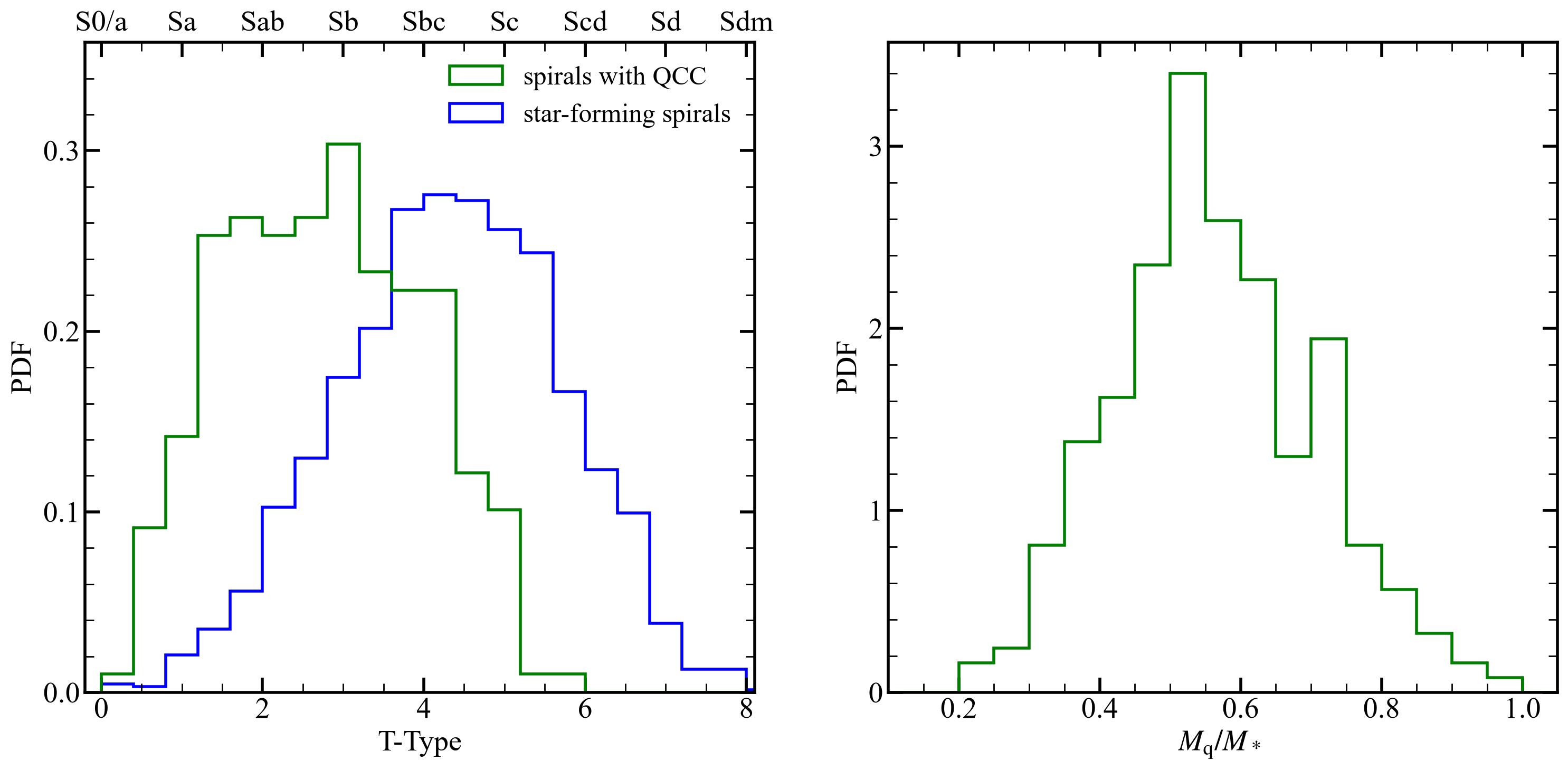}
    \caption{The probability density functions of T-Type (left) and the QCC mass ratio $M_{\rm q}/M_*$ (right). We additionally indicate the associated galaxy type along the top axis of the left panel \citep[see][]{dominguezsanchez_sdss-iv_2021}. Green histograms represent spirals with QCC, while blue histograms denote star-forming spirals (see Table \ref{tab:abbrev}).  }
    \label{fig:distribution}
\end{figure*}

As shown in Figure \ref{fig:bimodal_example}, some spiral galaxies have a quenched core in the central region. Previous studies adopted the bulge-disk decomposition technique to investigate the central component. Since we are particularly interested in the properties of the stellar population and star formation activity, we adopt the method shown in \cite{hongDynamicalHotnessStar2023} to identify the quenched central cores (QCCs). 
D4000 is an important indicator of stellar population and is strongly correlated with the SFR surface density (see, e.g., \citealt{Brinchmann_2004}). As a result, the D4000 map delivered by Pipe3D can provide valuable information on the spatial distribution of stellar population. 

Following \cite{hongDynamicalHotnessStar2023}, we adopt a threshold of $\rm D4000_{\rm th} = 1.55$ to distinguish quenched spaxels from star-forming ones. In addition to applying the Pipe3D quality mask, we restrict our analysis to spaxels with D4000 $\rm S/N>5$. A galaxy is classified as a fully quenched galaxy if at least 95\% of the spaxels within $1.5\,R_{\rm e}$ are quenched. Similarly, if at least 95\% of the spaxels within $1.5\,R_{\rm e}$ are star-forming, the galaxy is categorized as a fully star-forming galaxy. Galaxies with quenched spaxel fractions between 5\% and 95\% are defined as partially quenched galaxies. Among the spiral galaxies, applying these criteria yields 116 fully quenched spirals, 1,561 fully star-forming spirals, and 792 partially quenched spirals.

We identify QCCs only in the partially quenched spirals. For each spiral galaxy, we derive the quenched profile, $f_{\rm q}(<R)$, which is the fraction of quenched spaxels within a radius of $R$. All profiles presented in this work are corrected for inclination using the observed PA and ellipticity of the corresponding galaxy.  For galaxies with QCCs, the quenched profile usually decreases with increasing $R$. We define a minimum radius $R_{\rm q}$, within which $f_{\rm q}(<R_{\rm q})$ equals 95\%. We refer to the region inside $R_{\rm q}$ as the QCC. Throughout this analysis, we assume that the QCC has the same ellipticity as its host galaxy. Note that QCCs are not easily distinguished in edge-on galaxies. Therefore, we exclude galaxies with $b/a < 0.4$. For these QCCs, we derive the stellar mass $M_{\rm q}$ by summing the stellar mass of all spaxels within $R_{\rm q}$, and we estimate its uncertainty by assuming that its fractional error is the same as that of the total stellar mass. To ensure that the QCC is sufficiently well resolved, we further require $R_{\rm q} > 4$ arcseconds. We obtain 247 QCCs with $M_{\rm q}>10^{10}~\Msun$. If we relax the requirement to $R_{\rm q}>2$ arcseconds, we obtain 475 QCCs. Unless otherwise specified, our analysis focuses on QCCs with $R_{\rm q} > 4$ arcseconds. 

By definition, most of the regions within a QCC are quenched. Therefore, a QCC can be considered as a fully quenched system embedded in a spiral galaxy. It is thus interesting to compare this system with fully quenched galaxies, most of which are early-type galaxies. There are 1097 fully quenched ellipticals (hereafter quenched ellipticals) and 264 fully quenched S0s (hereafter quenched S0s) in our sample. It is also important to examine the properties of the outer disk (hereafter ODK) of a partially quenched spiral, defined as the part of the galaxy that remains after excluding its QCC. Note that spiral arms are expected to be the main feature of ODKs. We thus compare the ODKs with the fully star-forming spirals (hereafter star-forming spirals) in the following section. We list these galaxy/system samples in Table \ref{tab:abbrev}.

The left panel of Figure \ref{fig:distribution} presents the T-type distribution for spirals with QCCs. As shown there, roughly 70\% of these galaxies have T-type values greater than 2, indicating that they correspond to Sab or later-type spirals. For reference, we also present the T-type distribution for star-forming spirals. These spirals tend to exhibit a morphology corresponding to later Hubble types.
The right panel displays the distribution of the mass ratio ($M_{\rm q}/M_*$) between the QCCs and their host spirals. This distribution peaks near 0.5, indicating that, on average, QCCs and ODKs possess similar masses.

It is essential to compare the sizes of QCCs with those of other galaxies. We define the size of a QCC as the radius enclosing half of its stellar mass, i.e., $M_{\rm q}/2$. This radius is termed the half-mass radius ($R_{50,*}$). To ensure a consistent comparison, we also use the half-mass radius, rather than the effective radius, when characterizing galaxies. The central velocity dispersion of a QCC or a galaxy, $\sigma_{\rm c}$, is defined as the median value of the velocity dispersion within $0.2\,R_{50,*}$ for spaxels satisfying $\rm S/N>3$, with uncertainties estimated through spaxel resampling.

\section{Bimodality in spiral galaxies with QCCs}\label{sec_bimod}

\begin{figure*}[t]
    \centering
    \includegraphics[width=1\linewidth]{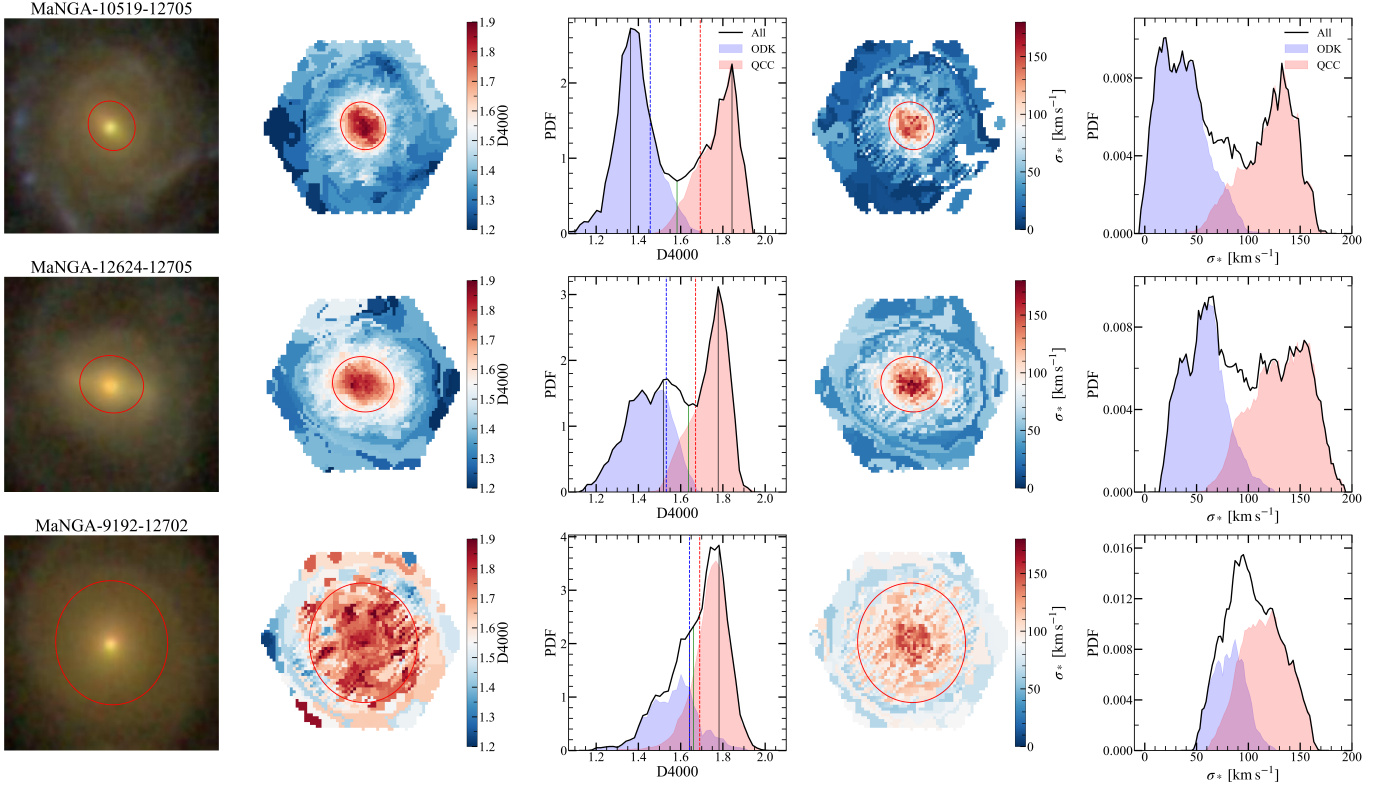}
    \caption{We show the representative spirals with strong (1st row), weak (2nd row), and no (3rd row) bimodality in D4000 and $\sigma_*$. The first-column panels show the SDSS images covering the same FOV as the Pipe3D datacubes. The second-column and fourth-column panels show the spatially resolved maps of the D4000 index and $\sigma_*$, respectively. The red elliptical curves in the these columns delineate the QCC regions. The third and fifth columns display the $\Sigma_*$-weighted probability distribution functions (PDFs) of $\rm D4000$ and $\sigma_*$ respectively. In both columns, the solid black curve represents the distribution of all available spaxels, while the blue and red shaded areas show the distributions for the ODK and QCC regions, respectively. In the third column, the vertical lines in blue, green, and red denote $\rm D_{\rm o80}$, $\rm D_{\rm v}$, $\rm D_{\rm c20}$, respectively, while the two vertical black lines mark the two identified peaks. See the text for the details.}
    \label{fig:bimodal_example}
\end{figure*}

QCCs are identified on the basis of the quenched fraction profiles of spaxels, which capture only the average radial behavior of D4000. It is therefore important to examine how the D4000 distribution in a QCC differs from that in the corresponding ODK, to which they are spatially connected. 
We visually examine the D4000 distributions of spaxels for each galaxy. As illustrated in Figure \ref{fig:bimodal_example}, some galaxies display a strongly bimodal distribution. In these cases, the large-D4000 peak is primarily contributed by the QCCs, while the small-D4000 peak is mainly associated with the ODKs. In contrast, for other galaxies, the D4000 distributions of QCCs and ODKs are heavily blended.
Therefore, it is important to develop a method to quantitatively characterize the bimodality linked to QCCs and ODKs.

To achieve this, for each partially quenched spiral, we first constructed the $\Sigma_*$-weighted D4000 distributions for the QCC and for the ODKs separately (see Figure \ref{fig:bimodal_example}). Throughout this paper, all D4000 distributions are smoothed using a top-hat filter with a width of 0.1.  
We then measured the $\rm 20th$ percentile of the QCC D4000 distribution, denoted as $\rm D_{c20}$, and the $\rm 80th$ percentile of the ODK D4000 distribution, denoted as $\rm D_{o80}$.  
When $\rm D_{c20}\le\rm D_{o80}$, the two distributions are expected to overlap significantly. Such systems are classified as galaxies without bimodality.  
Next, we considered the combined distribution, $h(\rm D4000)$, which includes both the QCC and ODK contributions (the black solid curves in Figure \ref{fig:bimodal_example}). For systems with $\rm D_{c20}>\rm D_{o80}$, we identified the maximum of $h$ in the region of $\rm D4000 < D_{o80}$, denoted by $h_{\rm op}$, and recorded its D4000 position as $\rm D_{op}$. Similarly, the maximum of $h$ in the region $\rm D4000 > D_{c20}$ is denoted by $h_{\rm cp}$, with its corresponding D4000 value of $\rm D_{cp}$. We then searched for the “valley,” defined as the minimum of $h(\rm D4000)$ between $\rm D_{c20}$ and $\rm D_{o80}$. The D4000 value and height at this minimum are denoted as $\rm D_v$ and $h_{\rm v}$, respectively. Examples are illustrated in Figure \ref{fig:bimodal_example}. 
If $h_{\rm op}\le h_{\rm v}$ or $h_{\rm cp}\le h_{\rm v}$, the system is again classified as a galaxy without bimodality. If both $h_{\rm op}\ge 1.4h_{\rm v}$ and $h_{\rm cp}\ge 1.4h_{\rm v}$ are satisfied, the system shows a pronounced bimodality in D4000 and is labeled as a strongly bimodal galaxy. The remaining systems display only mild bimodality and are therefore categorized as weakly bimodal galaxies. 

Figure \ref{fig:bimodal_example} shows illustrative examples of the bimodality. The left panels show the SDSS optical images. The middle-left panels display the corresponding D4000 maps, where the color scale is centered at $D4000=1.55$ (marked in white). This reference value is adopted to separate star-forming (blue) from quenched (red) regions (see Section \ref{subsec_QCCdef}). The middle panels present the D4000 distributions for the QCCs, the ODKs, and the entire galaxies. And the middle-right and right panels show the results for $\sigma_*$. Our visual inspection confirms that the method is robust in identifying bimodality. 
Our approach begins by separating QCCs and ODKs, instead of directly identifying peaks and valleys in the total distribution. This strategy has the advantage of explicitly linking the observed bimodality to these two structures and remains effective even when the total distribution contains three or more peaks.

We obtain 135 galaxies with strong D4000 bimodality, 57 with weak bimodality and 55 without significant bimodality. We also examine the distribution of stellar age and find bimodality in 157 galaxies. It is expected as D4000 is tightly correlated with the stellar age. We further applied this method to the distribution of stellar velocity dispersion ($\sigma_*$). Strong bimodality in the $\sigma_*$ distribution is detected in 119 galaxies and weak bimodality in 68 galaxies. Similarly to the bimodal D4000 distribution, the low-$\sigma_*$ peak is linked to the ODKs, while the high-$\sigma_*$ peak is linked to the QCCs. Approximately 62\% of spiral galaxies with QCCs show simultaneously D4000 and $\sigma_*$ bimodality. The fraction is comparable to those for the D4000 and $\sigma_*$ bimodality, which are 78\% and 76\%, respectively. Thus, the presence of $\sigma_*$ bimodality is strongly associated with the D4000 bimodality. Bimodality clearly emerges as a dominant characteristic of individual spiral galaxies hosting QCCs. Spiral galaxies that exhibit bimodality in both D4000 and $\sigma_*$ are referred to as spirals with bimodality, while the remaining spirals in which QCCs are detected are classified as spirals without bimodality (see Table \ref{tab:abbrev}).


\begin{figure*}[t]
    \centering
    \includegraphics[width=1\linewidth]{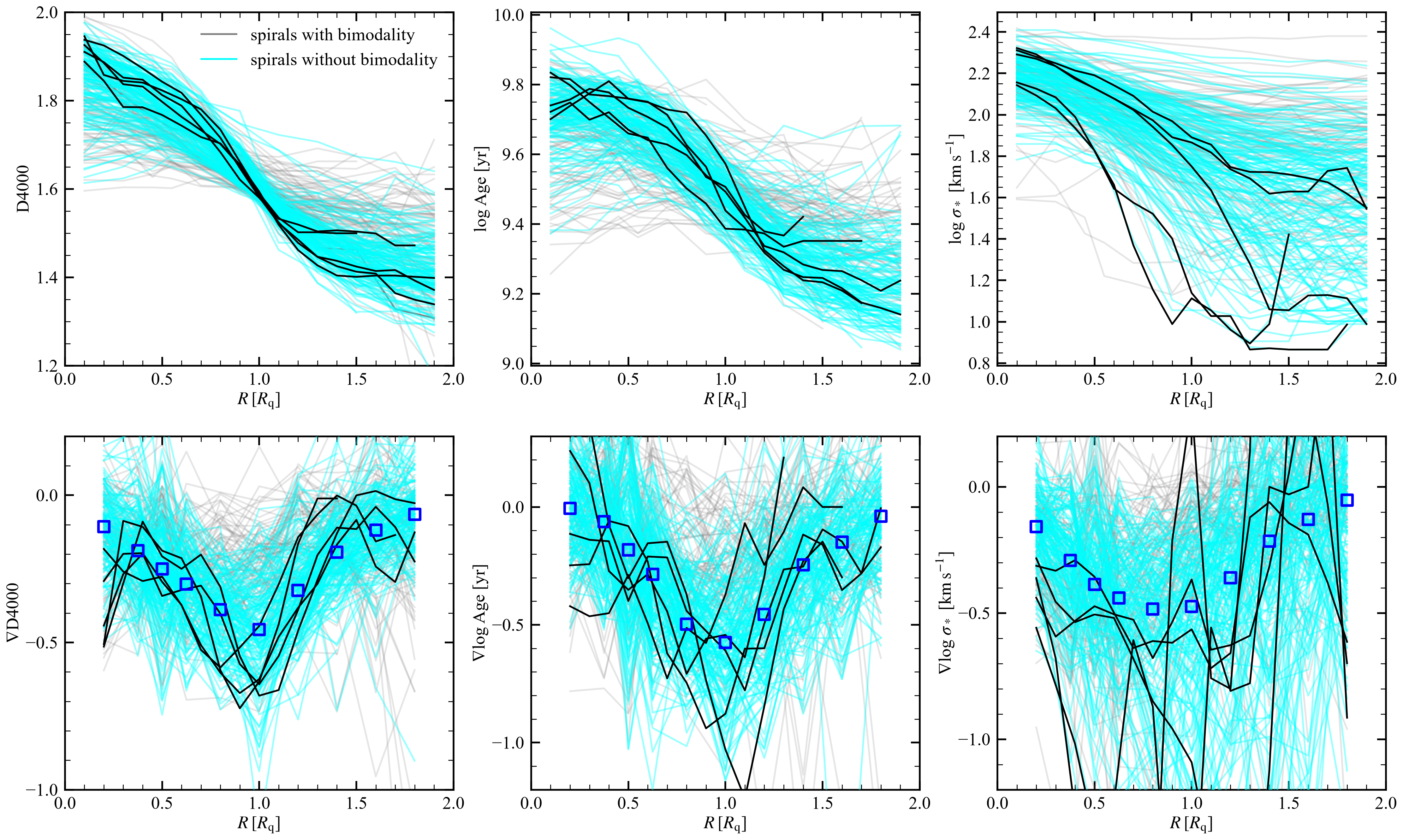}
    \caption{The upper panels show the D4000, stellar age and $\sigma_*$ profiles for the spiral galaxies hosting QCCs. The lower panels show the gradients of the three parameters as functions of radius. Here the radius is scaled with $R_{\rm q}$. The cyan ones show the results for the spirals with bimodality in both D4000 and $\sigma_*$, while the green ones show the results for the spirals without bimodality. The squares in the lower panels show the median profiles of the cyan ones. The black lines show the profiles for the five spirals with the largest $R_{\rm q}$ in arcseconds, which are expected to be the least affected by PSF. To derive the profiles of a galaxy, we first divided the spaxels from 0 to $2.0R_{\rm q}$ into 10 concentric elliptical annuli with a uniform width of $0.2 R_{\rm q}$, adopting its corresponding PA and ellipticity. To build the D4000 profile, we used only those spaxels with a reliable D4000 measurement ($\rm SNR > 5$), and likewise, only spaxels whose continuum spectra have $\rm SNR > 5$ were considered in deriving the age profiles, and only spaxels with spaxels with $\sigma_*$ $\rm SNR>3$ were used in deriving the $\sigma_*$ profiles. One possible reason for the large variation in the $\sigma_*$ gradient profiles is that the uncertainty in $\sigma_*$ is rather large. For the five spirals, we obtained the profiles using a smaller bin size of $0.1R_{\rm q}$.}
    \label{fig:gradprofile}
\end{figure*}


Figure \ref{fig:gradprofile} presents the D4000, stellar age, and $\sigma_*$ radial profiles for spiral galaxies that display bimodality in both D4000 and $\sigma_*$. All three quantities show pronounced declines with increasing radius. This behavior is expected, since in these systems QCCs have significantly higher D4000 (and thus older stellar ages) and larger $\sigma_*$ than their associated ODKs. Gradient profiles reveal more intriguing patterns. As the radius increases, the gradients initially become steeper and then flatten. This change in trend occurs near $R_{\rm q}$ and is visible in nearly all galaxies with bimodality, and for all three parameters. The large variation in the $\sigma_*$ gradient profiles is likely caused by the large uncertainty in $\sigma_*$ measurement.
However, the intrinsic gradient can be strongly smeared by the point spread function (PSF). To minimize this effect, we focus on the five galaxies with the largest $R_{\rm q}$ values in arcseconds, which are the least affected by the PSF. As anticipated, the gradients around $R_{\rm q}$ in these five galaxies are clearly steeper than those of the full sample. In particular, the drop and rise in the D4000 and age gradient profiles near $R_{\rm q}$ are sharp and narrow. We argue that the sharp change is an ubiquitous feature for spiral galaxies hosting QCCs, and that the smoother trend seen in the full sample is primarily caused by the PSF-induced smearing effect. The near-zero gradients at small (or large) radii indicate that the radial variations within QCCs (or within ODKs) themselves are relatively mild. The pronounced change in the gradients appears near the QCC boundary, demonstrating that QCCs are intrinsically different from ODKs.

\begin{figure*}[t]
    \centering
    \includegraphics[width=1\linewidth]{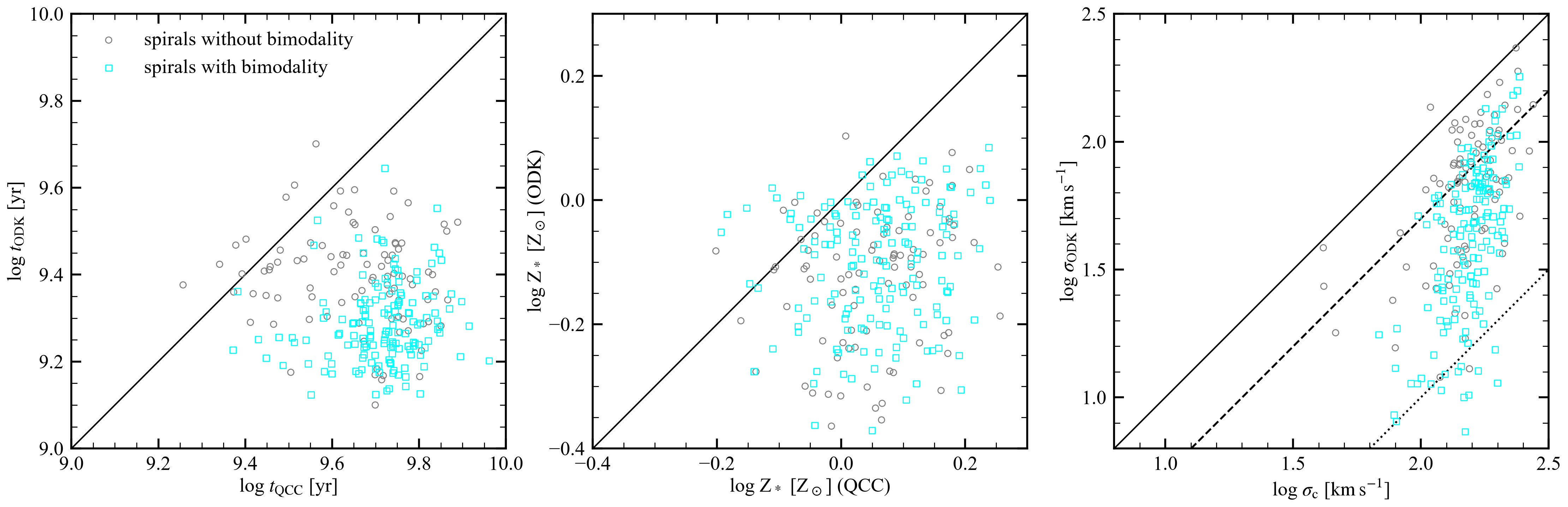}
    \caption{Comparison between QCC and ODK for spirals with QCCs. The spirals with bimodality in both D4000 and $\sigma_*$ are shown in cyan color, while the spirals without bimodality are shown in gray. From left to right, the panels display stellar age, metallicity, and the velocity dispersion. The black solid lines represent the 1:1 reference lines. The dashed and dotted line in the right panel represent the $\sigma_{\rm ODK}=\sigma_{\rm c}/2$ and $\sigma_{\rm ODK}=\sigma_{\rm c}/10$. }
    \label{fig:io_comp}
\end{figure*}

Figure \ref{fig:io_comp} compares the stellar ages, stellar metallicities, and stellar velocity dispersions of QCCs with those of ODKs. Because the properties of QCCs and ODKs in the vicinity of $R_{\rm q}$ are likely to be mutually contaminated, and because the internal radial variations within either QCCs or ODKs are relatively small, we characterize each QCC (or ODK) by taking the median of the spaxel properties within $R<0.2R_{\rm q}$ (or $1.4R_{\rm q}<R<1.6R_{\rm q}$). The only exception is that, for consistency, we use the central velocity dispersion $\sigma_{\rm c}$ (as defined in Section \ref{subsec_QCCdef}) to describe the dynamical state of a QCC.
The QCC ages ($t_{\rm QCC}$) lie between $10^{9.6}$ and $10^{9.8}$ years, while their outskirts are markedly younger, with ages ($t_{\rm ODK}$) between $10^{9.2}$ and $10^{9.4}$ years. It is consistent with previous studies based on bulge-disk decomposition (e.g. \citealt{JinY2024AA}). The age difference between the two components is quite large, suggesting that QCCs and ODKs form at different cosmic epochs. The stellar metallicities of the QCCs are all close to the solar value, while those of the ODKs are, on average, about 0.2 dex lower. There is no correlation between $t_{\rm QCC}$ and $t_{\rm ODK}$ or between the metallicities of these two components, implying that the formation of the two structures is not physically related. 
The ODKs are usually dynamically cold, compared to the QCCs. The coldest outskirts have $\sigma_{\rm ODK}<\sigma_{\rm c}/10$. About 74\% of galaxies have $\sigma_{\rm ODK}<\sigma_{\rm c}/2$. There exists a weak correlation between $\sigma_{\rm ODK}$ and $\sigma_{\rm c}$, which is mainly caused by the mass dependence.

Our findings indicate that QCCs and ODKs differ clearly in age, metallicity, and dynamical state. This sharp contrast leads to the bimodal distributions of D4000 and $\sigma_*$, as well as to the strongest gradient near the QCC boundary, all of which are commonly seen in many galaxies.
These results imply that the two components are assembled at different times and via distinct physical processes. In the next section, we carry out a detailed comparison of the properties of quenched early-type galaxies and QCCs, which may shed light on the formation pathway of QCCs.

\section{Similarity between QCCs and quenched early-type galaxies}\label{sec_s&d}

In this section, we present a systematic comparison between QCCs and quenched ellipticals/S0s. In particular, we compare them from several aspects: the size–mass relation, the $\sigma_{\rm c}$-$M_*$ relation, and stellar population and kinematic profiles. 
In addition, we provide a brief comparison between ODKs and star-forming spirals. This comparative analysis may offer key insights into the formation of QCCs and the evolution of their host galaxies.

\begin{figure*}[t]
    \centering
    \includegraphics[width=1\linewidth]{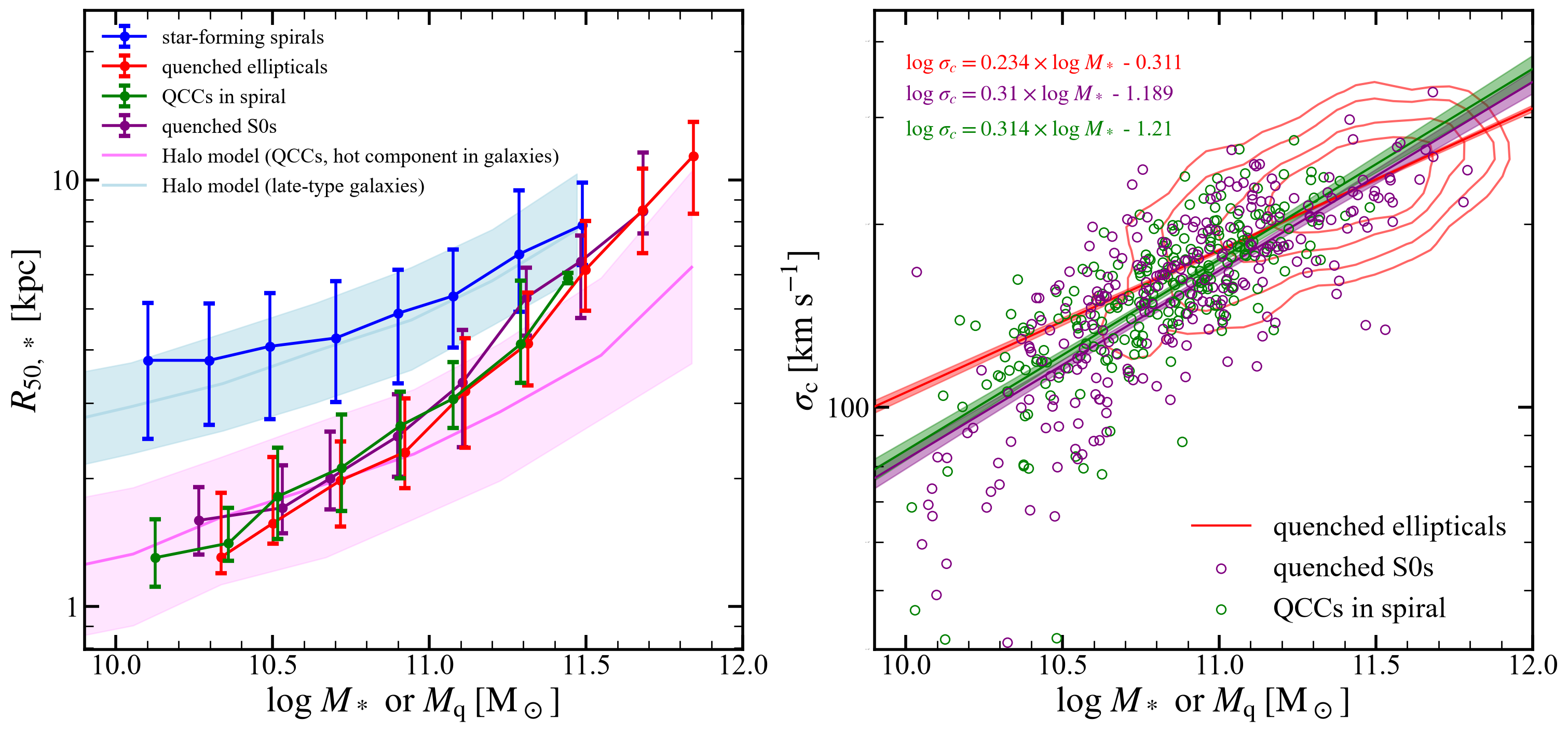}
    \caption{The size–mass and velocity dispersion-mass relations. In the left panel, the solid lines show the median $R_{50,*}$ for QCCs (green), star-forming spirals (blue), quenched ellipticals (red) and quenched S0s (purple). The error bars represent the 16th and 84th percentiles of the $R_{50,*}$ distribution. The shaded regions show the SMRs for modeled late-type galaxies (light blue) and QCCs (pink), taken from two-phase galaxy formation model\citep{Mo_2024, ChenY2025arXiv}. The right panel shows the $\sigma_\mathrm{c}$-$M$ relation. The green and purple open circles show the results for the QCCs in spirals and quenched S0s respectively, while the red contours show the results for quenched ellipticals. The red, purple, and green lines show the best fitting results for quenched ellipticals, quenched S0s, and QCCs, respectively. The shaded areas denote the uncertainties associated with these fitted results.
    }
    \label{fig:scalingR}
\end{figure*}

\subsection{Scaling relations}

We first examine the size–mass relation (SMR), a scaling relation widely used to trace evolutionary pathways throughout cosmic time (\citealt{Shen_2003}; \citealt{trujillo_luminositysize_2004}; \citealt{matharu_hst_2019}). It reflects the assembly history of galaxies and exhibits distinctly different behaviors for star-forming and quiescent systems, while also showing clear evolution with redshift (\citealt{Mosleh_2012}; \citealt{van_der_wel_3d-hstcandels_2014}; \citealt{Shibuya_2015}; \citealt{YangL_2025}; \citealt{Song2026A&A}). The SMRs for QCCs, star-forming spirals, and quenched ellipticals/S0s are shown in Figure \ref{fig:scalingR}. 
The SMR of star-forming spirals lies above that of quenched early-type galaxies and exhibits a flatter slope. 
Previous works reported similar differences between late-type and early-type galaxies (\citealt{Shen_2003}; \citealt{Lange_2015}). Our results match closely those presented in the literature, and the small discrepancies are likely attributable to the different selection criteria used for our sample. The quenched ellipticals and S0 galaxies exhibit identical SMRs, in agreement with an earlier study showing that the SMR difference between ellipticals and S0s is very small at low redshift\citep{Huertas-Company2013MNRAS}.
Intriguingly, the SMR of QCCs closely matches those of quenched early-type galaxies, and the scatter around the relations is likewise similar. 


The velocity dispersion-mass relation is another fundamental scaling relation and has been widely studied. This relation can also be used to study mass assembly history and understand quenching mechanisms (\citealt{Faber1976ApJ}; \citealt{Cappellari2013MNRAS}; \citealt{hongDynamicalHotnessStar2023}; \citealt{Mo_2024}). The right panel of Figure \ref{fig:scalingR} shows the $\sigma_{\rm c}$-$M_*$ relations for quenched ellipticals, S0s and QCCs. We use a linear relation,
\begin{equation}
    \log \sigma_{\rm c}=\alpha \, \log \, M_*+\beta,
    \label{equ:QGSR}
\end{equation}
to fit the three kinds of objects, respectively \citep[see][ for the details of the fitting method]{hongDynamicalHotnessStar2023}. The best fitting results are also presented. We have $\sigma_{\rm c}\propto M_*^{0.233\pm0.007}$ for ellipticals, $\sigma_{\rm c}\propto M_*^{0.310\pm0.016}$ for S0s and $\sigma_{\rm c}\propto M_*^{0.314\pm0.019}$ for QCCs.
Ellipticals have an apparently different slope than S0s and QCCs. On average, elliptical galaxies are more massive than the other two objects. The majority of ellipticals have $\log M_*/\Msun>11$, whereas the stellar masses of quenched S0s and QCCs span the range $10^{10}$–$10^{11.5}\,\Msun$.
Therefore, our result is consistent with early studies showing that the most massive galaxies have a different slope from other galaxies (e.g. \citealt{Cappellari2013MNRAS}). More interestingly, the $\sigma_{\rm c}$-$M_*$ relations for S0s and QCCs are very similar. This suggests that, in terms of their dynamical state, QCCs more closely resemble quenched S0s than ellipticals. We will come back to this later.

Similar to the QCCs examined in this work, bulges also occupy the central regions of galaxies. In contrast to QCCs, however, bulges are identified through the decomposition of galactic light profiles. Bulges are commonly separated into classical bulges and pseudobulges. Previous studies have shown that classical bulges follow scaling relations that are similar to those of early-type galaxies, whereas pseudobulges do not\citep{Kormendy2004ARA&A, Gadotti_2009, Kormendy2016ASSL}. This suggests that our QCCs are more likely associated with classical bulges. Nonetheless, there are also important differences between classical bulges and QCCs. For instance, some classical bulges are blue and are still forming stars\citep{Gadotti_2009, HuJ2024A&A}, whereas QCCs are completely quenched. Thus, classical bulges and QCCs cannot be considered strictly equivalent.


\subsection{Stellar Population}\label{subsec_spcomp}

\begin{figure*}[t]
    \centering
    \includegraphics[width=1\linewidth]{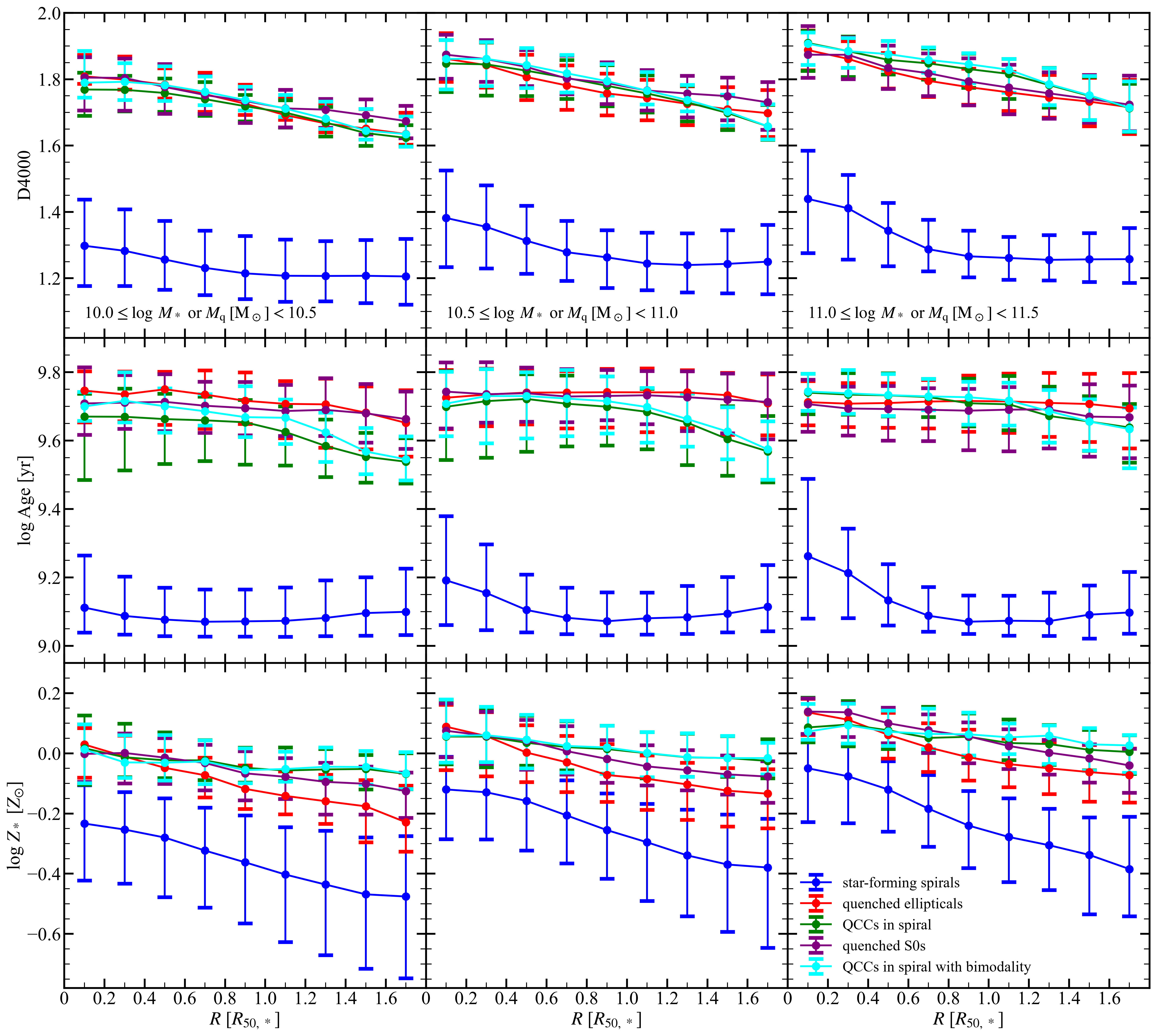}
    \caption{Radial profiles of stellar population characteristics for quenched ellipticals (red), star-forming spirals (blue), QCCs in spiral (green) and spirals with bimodality (cyan) in three mass bins as indicated in each panel. The lines show the median profiles, the error bars represent the $\rm 16th–84th$ percentiles of the distributions. The top, middle and bottom panels show the results for D4000, stellar age and stellar metallicity, respectively. We first derived the profiles for each galaxy, then obtained the median profiles and their dispersion of a sample. 
    To build the D4000 profile, we used only those spaxels with a reliable D4000 measurement ($\rm SNR > 5$), and likewise, only spaxels whose continuum spectra have $\rm SNR > 5$ were considered in deriving the stellar age and metallicity profiles.}
    \label{fig:sp_profile}
\end{figure*}

As a crucial indicator of stellar population properties, D4000 primarily arises from the cumulative metal-line absorption in the atmospheres of cool old stars. Consequently, D4000 is sensitive to the age and metallicity of stellar populations.  
The upper panels of Figure \ref{fig:sp_profile} present the radial D4000 profiles of QCCs, quenched ellipticals and S0s. For comparison, we also present the results for star-forming spirals. 
In Figure \ref{fig:sp_profile}, each line represents the median profile of the corresponding sample, and the error bars indicate the 16th and 84th percentiles of the distribution. We also present the stellar age and metallicity profiles in the same figure. 

In general, the radial profiles of QCCs closely resemble those of quenched ellipticals and S0s in amplitude, slope, and scatter across all three stellar mass bins. As stellar mass increases, both D4000 and metallicity rise in early-type galaxies and QCCs, whereas stellar age remains nearly constant. This indicates that, for all three types of systems, the mass dependence of D4000 is driven by the mass dependence of metallicity rather than by stellar age.  
In addition, early-type galaxies and QCCs both show negative radial gradients in D4000 and metallicity over the entire observed radial range. By contrast, the radial dependence of age is very weak at $R<1.2R_{50,*}$. As a result, the negative D4000 gradient is attributed to the metallicity gradient. Across all three mass bins, the metallicity profiles of S0s are slightly flatter than those of ellipticals, although the difference is small. Notably, the metallicity profiles of QCCs appear to be more similar to those of S0s than to those of ellipticals.

The most pronounced difference in the D4000 profiles between QCCs and early-type galaxies appears in the highest-mass bin, where QCCs exhibit D4000 values larger by about 0.05 at $R\sim R_{50,*}$. In contrast, the greatest difference in stellar age occurs in the lowest-mass bin. In this bin, QCCs are slightly younger than early-type galaxies across the entire radial range. We suspect that this may be caused by contamination from young stars. For galaxies that show bimodality, the old QCC component should be clearly separated from the young ODKs, so such contamination is expected to be minimal. We therefore present the results for these systems. As is evident, the new results agree much better with those of early-type galaxies than the original ones. Nevertheless, at large radii ($R>1.2R_{50,*}$), there remains a small discrepancy, likely still caused by contamination from the outskirts.

In contrast, the D4000 radial profiles of star-forming spirals are markedly lower than those of both QCCs and early-type galaxies in all three stellar mass bins. Their radial gradients also differ substantially, with steeper slopes at $R\le0.8R_{50,*}$ and nearly flat profiles at $R>0.8R_{50,*}$. Star-forming spirals host significantly younger and more metal-poor stellar populations than early-type galaxies, in agreement with previous studies (\citealt{Zheng_2017}; \citealt{Goddard_2017}). At large radii, the stellar age is about $10^{9.1}$ years and is nearly independent of stellar mass, while the metallicity increases with increasing $M_*$. The metallicity contrast between spiral and early-type galaxies shows only a mild dependence on radius.  
ODKs likewise contain younger and more metal-poor populations than QCCs (Figure \ref{fig:io_comp}). The median stellar age in ODKs is $10^{9.31}$ years, with the 16th and 84th percentiles at $10^{9.21}$ and $10^{9.44}$ years, respectively. Overall, the ODKs tend to be older than the outskirts of star-forming spirals.

\subsection{Kinematic properties}\label{subsec_skcomp}

\begin{figure*}[t]
    \centering
    \includegraphics[width=1\linewidth]{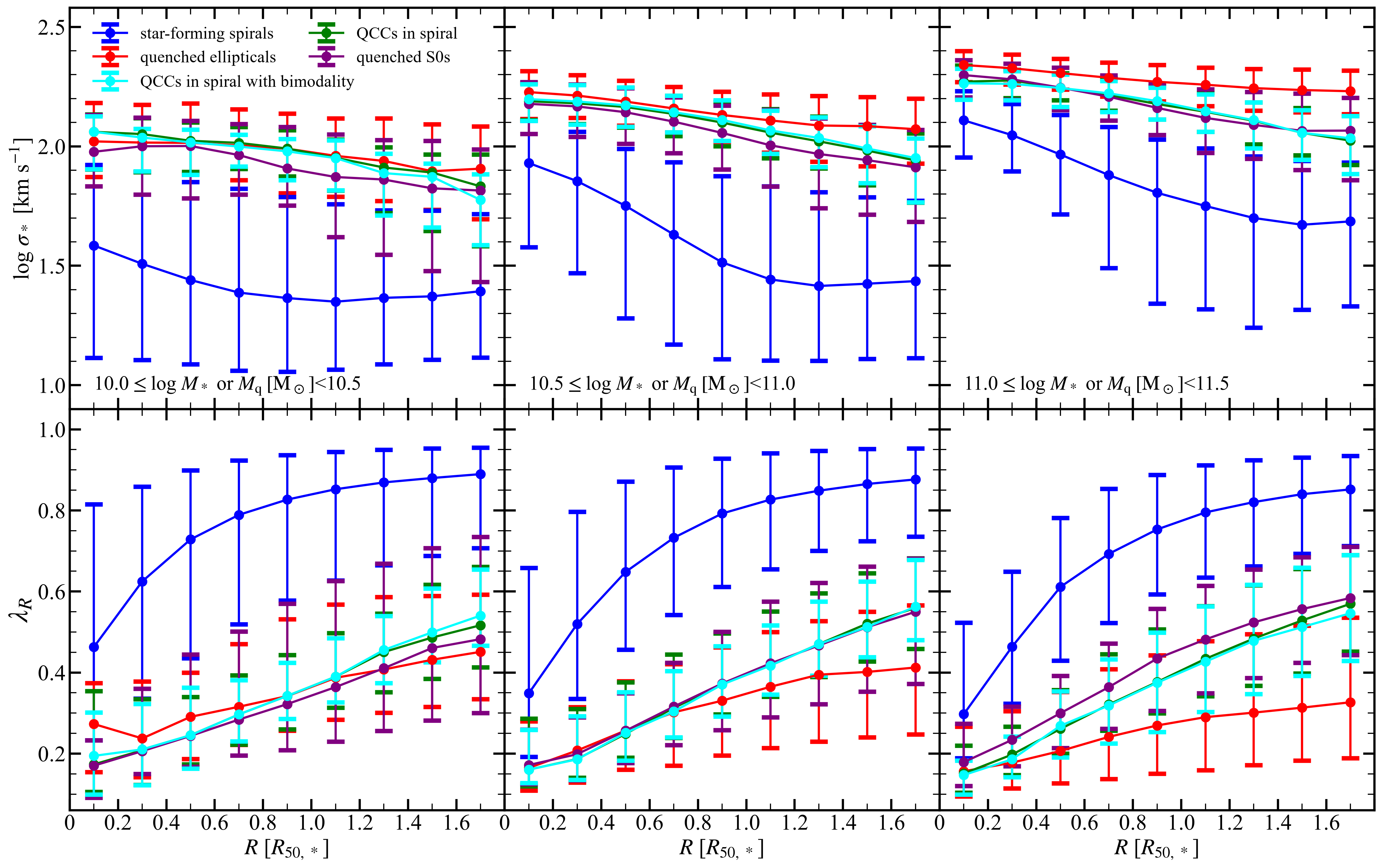}
    \caption{Radial profiles of stellar kinematic properties for quenched ellipticals (red), star-forming spirals (blue), QCCs in spiral (green) and spirals with bimodality (cyan) in three mass bins as indicated in each panel. The lines show the median results, the error bars represent the $\rm 16th–84th$ percentiles of the distributions. The top and bottom panels show the results for stellar velocity dispersion (only spaxels with $\rm SNR>3$ are used) and specific stellar angular momentum proxy (only galaxies with $\sin \theta_{\rm dyn}>0.2$ are included), respectively. }
    \label{fig:sk_profile}
\end{figure*}

The stellar velocity dispersion $\sigma_*$, which reflects the random motions of stars, is a key dynamical parameter that provides insight into the gravitational potential and formation history of galaxies. Figure \ref{fig:sk_profile} presents the median $\sigma_*$ as a function of radius for ellipticals, S0s, and QCCs. All three classes display nearly flat $\sigma_*$ profiles, but the differences between them, though small, are statistically significant. Quenched S0s show a slightly steeper decline, so their deviation from ellipticals increases toward larger radii. The difference between ellipticals and S0s also increases with increasing mass. Notably, QCCs also show a larger negative gradient, closely matching that of quenched S0s.

We then investigated the rotational properties of these systems. The spin parameter $\lambda_R$ is often used as a proxy for the specific stellar angular momentum (\citealt{emsellem_sauron_2007}; \citealt{Rong_2018}; \citealt{zhu_2023}).  Following \cite{emsellem_sauron_2007}, we define $\lambda_R$ as:
\begin{equation}
    \lambda_R=\frac{\langle R\vert V_i \vert \rangle}{\langle R \sqrt{V_i^2+\sigma_i^2} \rangle}=\frac{\sum_i F_i R_i V_i}{\sum_i F_i R_i \sqrt{V_i^2+\sigma_i^2}},
    \label{equ:lambdaR}
\end{equation}
where $F_i$, $V_i$, and $\sigma_i$ denote the flux, line-of-sight velocity, and velocity dispersion of the $i$-th spaxel, respectively, and $R_i$ is its distance to the center. In this work, $\lambda_{\rm R}$ is computed by summing over all spaxels within the corresponding ellipse with long axis of $R$. Therefore, $\lambda_{\rm R}$ characterize the mean rotation motion within $R$. 
Additionally, projection effects were corrected using dynamical inclination angles $\theta_{\rm dyn}$ from the data product of MaNGA DynPop project (\citealt{zhu_2023}). Galaxies with $\sin \theta_{\rm dyn} \leq0.2$ were excluded from the analysis to minimize potential errors introduced during the deprojection process. 
Note that $\lambda_{\rm R}=1$ indicates that stellar motions are entirely governed by ordered rotation, whereas $\lambda_{\rm R}=0$ indicates that stellar motions are fully dominated by random, disordered movements.

Figure \ref{fig:sk_profile} shows the median $\lambda_R$ as a function of radius for QCCs and quenched early-type galaxies. For both ellipticals and S0s, 
$\lambda_{\rm R}$ rises gradually with radius. However, the radial gradients for S0s are substantially steeper, and the difference between ellipticals and S0s becomes very pronounced at $R>1.2\,R_{\rm e}$ in the two highest mass bins. This behavior is expected given the presence of a disk component in S0s. In addition, the dependence of $\lambda_{\rm R}$ on $M_*$ differs between S0s and ellipticals. For ellipticals, the spin declines markedly with increasing $M_*$, whereas for S0s it increases slightly with $M_*$. Taken together, these trends indicate distinct formation pathways and evolutionary histories for the two galaxy classes. The most striking result is that QCCs exhibit median $\lambda_{\rm R}$ profiles that are very similar to those of quenched S0s in all three mass bins. This agrees with the behavior of $\sigma_*$. Evidently, both QCCs and S0s, on average, rotate more rapidly than ellipticals of comparable stellar mass. The quantity $\lambda_{R_{\rm e}}$, measured within $R_{\rm e}$, is commonly used to distinguish fast and slow rotators (e.g., \citealt{Cappellari_2016}; \citealt{graham_sdss-iv_2018}; \citealt{zhu_2024}).
For instance, galaxies with $\lambda_{R_{\rm e}}<0.4$ ($\ge 0.4$) are classified as slow (fast) rotators (\citealt{Wang&Peng_2025}). We adopted a similar parameter, $\lambda_{R_{50,*}}$, and found that the fraction of slow rotators among quenched ellipticals rises from 58.3\% at $10<\log M_*/\Msun<10.5$ to 80.9\% at $11.0<\log M_*/\Msun<11.5$, while for QCCs (S0s) the fraction decreases from 69.2\% (59.5\%) to 53.2\% (38.1\%). 


Star-forming spirals exhibit $\sigma_*$ and $\lambda_{\rm R}$ profiles that differ substantially from those of quenched systems. Their $\sigma_*$ ($\lambda_{\rm R}$) profiles are markedly lower (higher) than the corresponding profiles of ellipiticals and S0s. Notably, the $\sigma_*$ profiles of star-forming spirals show a much stronger dependence on $M_*$ than those of quenched systems, whereas their $\lambda_{\rm R}$ profiles are almost insensitive to $M_*$. The radial behavior in star-forming spirals also contrasts sharply with that of the quenched systems. For example, their $\lambda_{\rm R}$ rises steeply with radius in the inner regions and then levels off at $\lambda_{\rm R}\sim0.85$ for $R\ge R_{50,*}$. ODKs, which are spatially connected with QCCs, are also much dynamically colder than QCCs. For each ODK, we derive a single $\lambda_{\rm R}$ using all available spaxels. The median, 16th, and 84th percentiles of the ODK $\lambda_{\rm R}$ distribution are 0.91, 0.78, and 0.97, respectively. These values are comparable to those of the outskirts of star-forming spirals, indicating that their ODKs are dominated by ordered rotation. This is consistent with the fact that the ODKs are disk-like structures.

\section{Summary and discussion}\label{sec_sum}

In this study, we carry out a detailed investigation of quenched central cores (QCCs) in spiral galaxies using IFU observations from the MaNGA survey. QCCs are located at the centers of spiral galaxies and are identified solely from the spatial distribution of the D4000 index. Compared to the outer disks (ODKs) of these galaxies, the QCCs are markedly older, more metal-enriched, and dynamically hotter (Figure \ref{fig:io_comp}). Due to the pronounced contrast in stellar populations and dynamical status between ODKs and QCCs, more than 60\% of spirals with QCCs display bimodal D4000 and $\sigma_*$ distributions, where the low-D4000/$\sigma_*$ peaks trace ODKs and the high-D4000/$\sigma_*$ peaks trace QCCs (Figure \ref{fig:bimodal_example}). We also observe abrupt variations, a decline followed by an increase, in the gradient profiles of D4000 and stellar age around the QCC boundary (Figure \ref{fig:gradprofile}). This pattern suggests that these galaxies experienced a non-continuous assembly history.

To explore the origin of QCCs, we compare them with quenched early-type galaxies, including both ellipticals and S0s. Both QCCs and quenched early-type galaxies are fully quenched systems, meaning that more than 95\% of their spaxels are quenched. We find that QCCs follow the same size–mass and velocity–dispersion–mass relations as quenched early-type galaxies (Figure \ref{fig:scalingR}). QCCs and quenched early-type galaxies also show similar radial profiles of stellar age and metallicity, as well as comparable dynamical structures, as indicated by their velocity dispersion and spin parameter profiles (Figures \ref{fig:sp_profile} and \ref{fig:sk_profile}). In both populations, the stellar content is dominated by old, metal-rich stars, and the systems are largely dynamically hot. When we distinguish between ellipticals and S0s, we find that QCCs resemble quenched S0s more closely than ellipticals in their dynamical properties. The velocity–dispersion–mass relation of QCCs has a slope that is nearly identical to that of S0s but significantly different from that of ellipticals. The velocity dispersion and spin parameter profiles of QCCs are also very similar to those of S0s and clearly distinct from those of ellipticals. On average, QCCs and S0s exhibit higher rotation than ellipticals.
By contrast, the ODKs are primarily composed of young stars and supported by rotation, much like star-forming spiral galaxies.

These results point to a scenario in which QCCs and quenched early-type galaxies are produced through the same initial formation channels but follow distinct evolutionary paths afterward. Their similarities in structure, stellar populations, and dynamical properties strongly imply that QCCs and early-type galaxies formed at the same cosmic time, underwent comparable mass assembly histories, and were quenched by the same physical mechanisms. These systems originated through violent events, such as major mergers (\citealt{Naab_2014}; \citealt{Penoyre_2017}; \citealt{Legos_2018}; \citealt{Rong_2025}) or rapid mass build-up in the early universe (see, e.g., \citealt{Mo_2024}). After this phase, however, the mechanisms acting on these galaxies begin to diverge. In some galaxies, heating of the circumgalactic medium (CGM) efficiently counteracts cooling, allowing them to remain quiescent and dynamically hot up to the present epoch: these systems are what we identify as quenched early-type galaxies. In other galaxies, on the contrary, heating becomes ineffective, enabling the rejuvenation of star formation. This rejuvenation phase is much gentler than the preceding violent stage. The cold gas gradually spirals inward and settles into stable, cold disks. Galaxies in this latter category are now classified as spiral galaxies with QCCs embedded at their centers. Therefore, our findings suggest that some galaxies transform from early-type/lenticular galaxies to spiral galaxies, in contrast to the conventional view of how galaxies evolve.


\begin{figure}[t]
    \centering
    \includegraphics[width=1\linewidth]{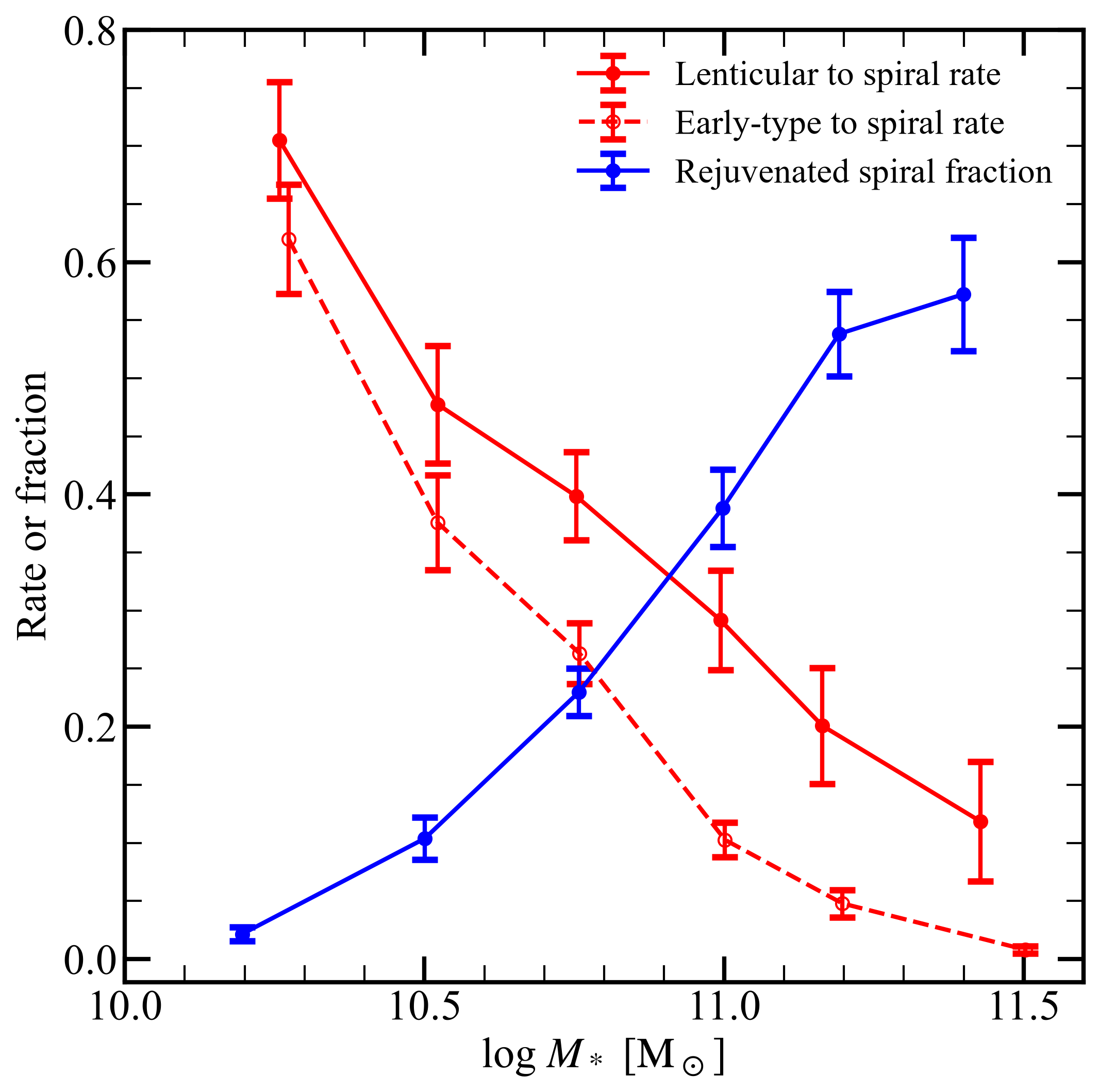}
    \caption{The red lines show the morphological transformation rates from quenched lenticular galaxies ($f_{\rm L2S}$, solid) or quenched early-type galaxies ($f_{\rm E2S}$, dashed) as a function of the progenitor stellar mass. The progenitor mass of a QCC is set to $M_{\rm q}$, while the progenitor mass of an elliptical or S0 galaxy is set to its stellar mass. The blue line shows the fraction of rejuvenated spiral galaxies among all spiral galaxies as a function of stellar mass. To compute the rates and fraction, we use QCCs with $R_{\rm q}\ge2$ arcseconds rather than those with $R_{\rm q}\ge4$ arcseconds. To correct for the selection effect for MaNGA galaxies, we adopt the \textsc{ESRWEIGHT} weights described in \cite{Wake_2017}. The error bars are estimated using the bootstrap technique.}
    \label{fig:rej_frac}
\end{figure}

\cite{Mo_2024} proposed a two-phase framework for galaxy formation. In this picture, dynamically hot systems, namely QCCs and quenched early-type galaxies, form rapidly during the early, fast-accretion phase of dark matter halos, whereas dynamically cold systems, including ODKs and star-forming spiral galaxies, build up more gradually throughout the subsequent slow-accretion phase of these halos. Following \cite{Mo1998MNRAS} and \cite{ChenY2024MNRAS}, we use a halo-based empirical relation to link the size of a late-type and that of a QCC (or an early-type galaxy) with their host halos via
\begin{equation}
    R_{\rm 50, late}=f_{\rm late} R_{\rm h}, \label{eq:Rlate}
\end{equation}
\begin{equation}
    R_{\rm 50, early}=f_{\rm early} R_{\rm f}, \label{eq:Rearly}
\end{equation}
respectively. Here $f_{\rm late}$ and $f_{\rm early}$ are free parameters, $R_{\rm h}$ is the size of the host halo at the present epoch, and $R_{\rm f}$ is the size of the host halo at redshift $z_{\rm f}$, the epoch at which the fast accretion of the halo ceases so that the dynamically hot component of the galaxy stops growing. We adopt the sample of central galaxies at $z=0$ from the semi-analytic model of \cite{ChenY2025arXiv}, which provides the stellar masses of the dynamically hot and cold components of each galaxy, as well as $R_{\rm h}$ and $R_{\rm f}$ of its host halo. We select late-type galaxies from the sample as those with a disk mass fraction larger than 0.5, and QCCs as the hot components of all galaxies. We use Eqs. \ref{eq:Rlate} and \ref{eq:Rearly} to calculate the sizes of late-type galaxies and QCCs, respectively, and calibrate $f_{\rm late}$ and $f_{\rm early}$ to match our observed SMRs. We find that $f_{\rm late}=0.02$ and $f_{\rm early}=0.01$ provide reasonable matches, as shown by the shaded bands in Figure \ref{fig:scalingR}. The value of $f_{\rm late}$ so obtained is consistent with previous observational results\citep[e.g.][]{Kravtsov2013ApJ, Huang2017ApJ, Mishra2023MNRAS}. 
More importantly, the small physical sizes of halos at early epochs (i.e. $z_{\rm f}$) naturally explain the offset of QCC sizes from those of late-type galaxies at given $M_*$, providing further support for the idea that QCCs are products of early galaxy formation before their ODKs could prevail. 

An alternative interpretation is that no substantial morphological transformation has taken place throughout the evolutionary history of these galaxies. In this picture, the progenitors of the spiral galaxies under consideration were already spiral systems from the outset. Subsequent processes then acted to quench star formation in their central regions and to dynamically heat the stellar components, thereby producing the QCCs, while the outer regions preserved a dynamically cold, disk-like structure and continued to sustain star formation. This evolutionary scenario faces two main challenges. First, it has difficulty accounting for the rapid, nearly discontinuous changes observed in the radial gradients of physical properties at the QCC boundary, as a continuous evolution would be expected to generate more gradual variations in these gradients. Second, it cannot readily explain why the structural and stellar population properties, as well as the scaling relations of QCCs, closely resemble those of quenched early-type galaxies. Such a high degree of similarity strongly suggests that these two classes of systems share a common evolutionary pathway and have been subject to analogous physical processes. In contrast, within this scenario, the evolutionary histories of the spirals with QCCs and those of early-type galaxies are intrinsically different, in tension with the observational evidence. Consequently, if these galaxies had indeed preserved a spiral morphology throughout their evolution, our results would impose very stringent constraints on the nature and efficiency of the underlying physical mechanisms.

A key question is how frequently the early-type galaxies experience morphological transformation to spiral galaxies during their subsequent evolution. Suppose that the progenitors of QCCs were dominated by S0 galaxies, the transformation rate from lenticular galaxies to spiral galaxies, $f_{\rm L2S}$, can be computed as the ratio of the number of QCCs to the total number of QCCs plus quenched S0s. If ellipticals and S0 galaxies are both able to evolve into spirals, then the transformation rate from early-type to spiral systems, $f_{\rm E2S}$, can be evaluated as the number ratio of QCCs to QCCs plus quenched early-type galaxies. It is also essential to present the fraction of rejuvenated galaxies within the spiral population, defined as the number ratio of QCCs to the total spiral galaxy sample.

Figure \ref{fig:rej_frac} presents the rates and fraction as a function of stellar mass. As shown in Section \ref{subsec_QCCdef}, QCCs are not identified in $\sim69$\% of partially quenched spirals. An important reason is that some QCCs have $R_{\rm q}<4$ arcseconds and thus fail our selection threshold. To account for this, we relax the requirement to $R_{\rm q}\ge2$ arcseconds to calculate the transformation rate. As illustrated, $f_{\rm L2S}$ can reach $\sim70$\% at $\log M_*/\Msun\sim10.2$, and gradually decreases to $\sim40$\% at $\log M_*/\Msun\sim10.7$. At $\log M_*/\Msun>11.2$, the fraction becomes less than 20\%. Due to the inclusion of ellipticals, which are more massive than S0s, $f_{\rm E2S}$ decreases more dramatically. However, the rate is still pronounced. We also present the fraction of rejuvenated galaxies among the spirals.  At $\log M_*/\Msun\sim10.7$, about 20\% of spirals are galaxies rejuvenated from early-type galaxies. This fraction increases to about 55\% for the most massive spirals ($\log M_*/\Msun>11.2$). The intrinsic fractions are higher, since some QCCs remain undetected owing to the limited spatial resolution or being overwhelmed by the star formation regions. 
Thus, our analysis suggests that the transformation from early-type galaxies to spiral morphology is a common evolutionary pathway in our Universe.

Our results are crucial for investigating how star formation is quenched in galaxies. Quenching can proceed through two different mechanisms (see e.g. \citealt{Croton_2006}). The first is the depletion and/or ejection of the cold gas inside galaxies, which ultimately shuts down star formation. This mechanism appears to have worked in both QCCs and quenched early-type galaxies.
The second mechanism suppresses the cooling of the CGM, so that galaxies are deprived of new cold gas and can remain quiescent. This mode is efficient in quenched early-type galaxies but inefficient in QCCs, allowing ODKs, young and cold disks, to gradually build up around the QCCs.

Therefore, both quenched galaxies and QCCs are essential for probing the physical processes that terminate star formation. In most previous works, quiescent and star-forming systems are distinguished using a threshold in specific star formation rate (sSFR), typically $10^{-11}\,\rm year^{-1}$. With this criterion, 32\% of spirals with QCCs are classified as star-forming, while the rest are labeled as quiescent. This standard scheme complicates the interpretation of results, because some galaxies tagged as star-forming have already undergone quenching events, and some galaxies identified as quiescent are, in fact, experiencing gas accretion and rejuvenation. Green valley galaxies lie between star-forming and quiescent galaxies, making them a key population for investigating how galaxy quenching occurs. \cite{Salim_2014} defined green valley galaxies as those with $-11.8<\log sSFR<-10.8$. In our sample, 191 galaxies (77\%) satisfy this definition. Green valley galaxies are commonly viewed as systems in transition from active star formation to quiescence. Our analysis suggests that this is misleading, since a subset of these objects is more plausibly moving from a quiescent state back toward a star-forming phase. Thus, a joint analysis of quenched galaxies and QCCs can provide a more comprehensive view of the underlying physical mechanisms.

The distinction between quenched early-type galaxies and QCCs arises from their divergent evolutionary tracks during the second phase. Unlike quenched galaxies, QCCs acquire a substantial amount of cold gas at this stage. Mergers can deliver cold gas to QCCs (\citealt{George_2017}; \citealt{Rathore_2022}; \citealt{WangYj_2025}). However, the cold gas supplied by mergers is usually insufficient to explain the $M_{\rm q}/M_*$ distribution (Figure \ref{fig:distribution}), which has a median of 0.55 and ranges from 0.41 (16th percentile) to 0.71 (84th percentile). In addition, major mergers are generally thought to destroy disks rather than assemble them, inconsistent with the morphology distribution (Figure \ref{fig:distribution}). Therefore, mergers only contribute to the rejuvenation of a fraction of galaxies where rejuvenation is weak. Radio AGN feedback can heat the CGM, inhibit gas cooling, and is widely considered the main “maintenance-mode” mechanism. Recently, \cite{Liu2025_AIRadio} argued that radio AGN feedback is only effective in a subset of massive quiescent galaxies. If radio AGN feedback is inefficient in some galaxies, gas can cool and fuel new star formation, allowing these quenched systems to eventually evolve into QCCs embedded in spiral galaxies at the present epoch. In contrast, for systems with efficient radio feedback, galaxies remain quiescent, undergo little subsequent evolution, and ultimately appear as quenched early-type galaxies in the local universe. Our investigation can therefore provide a stringent constraint on the efficiency of maintenance-mode feedback.

\section*{Acknowledgements}

This work is supported by the National Natural Science Foundation of China (NSFC, Nos. 12192224, 12595312). HYW thanks the support of CAS Project for Young Scientists in Basic Research, Grant No. YSBR-062, the New Cornerstone Science Foundation through the XPLORER PRIZE, and the China Manned Space Program
with grant no. CMS-CSST-2025-A04. 
The authors gratefully acknowledge the support of Cyrus Chun Ying Tang Foundations. 
The work is supported by the Supercomputer Center of University of Science and Technology of China.

\bibliography{ref.bib}

\nolinenumbers
\clearpage

\end{document}